# Visualizing Anderson Localization of Ultracold atoms in a disordered potential using Monte Carlo method

S.Datta

Department of Theoretical Physics

Indian Association for the Cultivation of Science

2A & 2B Raja S. C. Mullick Road,

Kolkata 700 032, India

We numerically study the effect of Anderson localization on a trapped Bose-Einstein condensate in 1d and 3d in a disordered potential. Simulations are performed in continuous space using canonical ensemble. Owing to the high degree of control over the system parameters we specifically study the interplay of disorder and interaction in the system. Our simulations clearly demonstrates the previously established fact that the Anderson localization in quasi-periodic system mimics the transition from extended to localized state in disorder induced interacting system in higher dimension. We also numerically show that strong disorder can be introduced in ultracold gases in 3d which gives rise to reasonably short localization length. We observe that as the interaction strength increases, the wave functions become more and more delocalized. For probing the localization we calculate the mean square displacements as an alternative for finding the Lyapunov exponents.

**1 Introduction:**

Anderson localization [1] (AL) corresponds to a quantum interference of the multiply scattered non-interacting particles from the random potentials and is characterized by the exponential decay of wave-function over a typical distance, the localization length. This localization denies the possibility of ordinary conduction and is not at all obvious. AL was predicted fifty years ago in the context of the study of conductance of electrons in crystals. However, It has not been observed for matter waves due to the complexity in the inherent structure of solids and the complex interaction associated with it. Fortunately the recent emergence of the field of ultracold physics offers new approaches for these issues as the parameters involved in the interaction are so much controllable that ultracold atoms can realize quantum simulators i.e., they provide platforms to investigate various fundamental models which were hitherto inaccessible. The combination of ultracold atoms and the optical potential usually provides a platform to study disorder and interaction. A disordered potential can be realized either with laser speckles or quasi-periodic potentials and AL in 1d has been recently observed for both the disorder [2,3]. Recently localization of ultracold atoms in 3d ( both in bosonic and fermionic system) has been observed independently by Institut d'Optique[4] and Urbana-Champaign group[5].A pedagogical and a very comprehensive theoretical development has been provided in the recent reviews by B Shapiro[6]. It is known that system in 1d and 2d the eigenstates are localized for all values of disorder strength[7]. In 3d there exists an energy known as 'mobility edge' which separates the localized and extended states. Self consistent theory of localization enables one to calculate the localization length and mobility edge in 3d. But these are not fully exact. All of these issues call for investigations of AL in higher dimensions. Since Gross- Pitaevski or other mean field techniques break down for strong interaction, we propose to study the effect of Anderson localization in cold atoms with Quantum Monte Carlo simulation based on Feynman-Kac path integration method[8]. Specifically we simulate two cases namely (1) 1d bichromatic lattices and (2)ultracold gas in 3d( in both cases we use ultracold gases made up of $Rb^{87}$ atoms). We simulate bichromatic lattices in 1d as it mimics the Aubry-André (A-A) [9] model which shows an interesting feature of transition from extended to exponentially decayed states. In Section 1 we have included the introduction. In Section 2 we have described the model used in the simulation work, unit analysis and precision in the numerical scheme used in the simulations. Section 3 contains all the localization plots related to 1d and 3d AL. Section 4 demonstrates the delocalization induced by interaction in various disorder in 1d and 3d. In section 5 we analyze the results of the various simulations.

## 2 The model:

### 2.1 Theory

We study the effect of Anderson localization of a d dimensional system of N $Rb^{87}$ atoms with a disordered potential and a short ranged interaction by Feynman-Kac path integral technique. Let us consider a Bose-Einstein condensate of N ( $N$ >>1) atoms of mass m trapped in d dimensional spherically symmetric harmonic potential $V_{trap}$, a random potential $V_R$ and a repulsive interaction $V_{int}$. In the Schrödinger picture the random Hamiltonian then can be represented as

$$H = -\Delta/2 + V \quad \text{with} \quad V = V_{trap} + V_{int} + V_R$$

Mostly at T=0, Gross-Pitaevski (GP) technique is applied for the calculation of energy, density of a many body system. The mean field approach such as GP for solving many body dynamics is only approximate. We propose to apply Diffusion Monte Carlo technique known as Feynman-Kac path integral simulation to the above quantum gas. To connect Feynman-Kac ( FK ) or Generalized Feynman-Kac ( GFK) to other many body techniques our numerical procedure has a straight-forward implementation[10] to Schrödinger's wave mechanics. For the Hamiltonian $H = -\Delta/2 + V(x)$ consider the initial-value problem

$$\frac{d\psi(x,t)}{dt} = \left(\frac{\Delta}{2} - V(x)\right)\psi(x,t), \tag{1}$$

with $x \in R^d$, $d = 3N$ and $\psi(x,0) = 1$. The Feynman–Kac solution to this equation is

$$\psi(x,t) = E\exp[-\int_0^t V(x(s))ds], \tag{2}$$

where x(t) is a Brownian motion trajectory and E is the average value of the exponential term with respect to these trajectories. The lowest energy eigenvalue for a given symmetry can be obtained from the large deviation principle of Donsker and Varadhan[11]

$$\lambda = -\lim_{t\to\infty}\frac{1}{t}\ln E\exp[-\int_0^t V(x(s))ds], \tag{3}$$

The above formula is not limited to calculating only the ground state but it provides the lowest exited states as well.

In dimension higher than 2, the trajectory x(t) escapes to infinity with probability 1. As a result, the important regions of the potential are sampled less and less frequently and the above equation converges slowly. Now for this analysis we do not incorporate importance sampling as the trial function chosen would impose some extraneous localization in the system and this localization might overshadow the localization induced by the disorder. We merely employ the bare Feynman-Kac path integral technique here.

Numerical work with bare Feynman-Kac procedure employing modern computers was reported [12] for the first time for few electron systems after forty years of the original work [8(b)] and seemed to be really useful for calculating atomic ground state. A fairly good success in atomic physics motivated us to apply it Condensed matter Physics[13]. Here in this paper we apply Feynman-Kac procedure to Bose-Einstein Condensate of $Rb^{87}$ to visualize Anderson localization in cold atoms.The normalized version of the many body density in the coordinate representation can be represented by [14]

$$\rho(x_1, x_2, \ldots \ldots x_{N:} x_1', x_2', \ldots \ldots x_N' = \sum_i w_i \ \rho(x_1, x_2, \ldots \ldots x_{N:} x_1', x_2', \ldots \ldots x_N')$$

where $w_i = \frac{e^{-\beta E_i}}{\sum_i e^{-\beta E_i}}$

$$\rho(x_1, x_2, \ldots \ldots x_{N:} x_1', x_2', \ldots \ldots x_N';\beta) = \frac{\sum_i e^{-\beta E_i}\psi_i(x_1 \ldots\ldots x_N)\psi_i^*(x_1' \ldots\ldots x_N')}{\sum_i e^{-\beta E_i}}$$

For $\beta \to \infty$ , only the ground state contributes or in other words

$\sum_i e^{-\beta E_i}\psi_i(x_1 \ldots \ldots x_N)\psi_i^*(x_1' \ldots \ldots x_N') \to e^{-\beta E_0}\psi_i(x_1 \ldots \ldots x_N)\psi_i^*(x_1' \ldots \ldots x_N')$ and $\sum_i e^{-\beta E_i} \to e^{-\beta E_0}$

Then the many body density becomes

$$\rho(x_1, x_2, \ldots \ldots x_{N:} x_1', x_2', \ldots \ldots x_N') = \frac{e^{-\beta E_0}\psi_i(x_1 \ldots \ldots x_N)\psi_i^*(x_1' \ldots \ldots x_N')}{e^{-\beta E_0}}$$

$$\rho(x_1, x_2, \ldots \ldots x_{N:} x_1', x_2', \ldots \ldots x_N') = \psi_i(x_1 \ldots \ldots x_N)\psi_i^*(x_1' \ldots \ldots x_N')$$

For diagonal density function becomes

$$\rho(x_1, x_2, \ldots \ldots x_N; x_1, x_2, \ldots \ldots x_N) = \psi_i^2(x_1, x_2, \ldots \ldots x_N)$$

In terms of Feynman-Kac solution the density function becomes

$$\rho = \left| E \exp[-\int_0^t V(x(s))ds] \right|^2$$

## 2.2 Numerical details and the precision of the calculations.

### 2.2.1 Numerical details:

The formalism given in the Section 1 is valid for any arbitrary dimensions d( for a system of N particles in three dimensions $d = 3N$). Generalizations of the class of potential functions for which Eqns 2 nad 3 are valid are given by Simon [15] and include most physically interesting potentials, positive or negative, including, in particular, potentials with $1/x$ singularities. It can be argued that the functions determined by Eq(3) will be the one with the lowest energy of all possible functions independent of symmetry. Although other interpretations are interesting, the simplest is that the Brownian motion distribution is just a useful mathematical construction which allows one to extract the other physically relevant quantities like density, mean square displacement along with the ground and the excited state energy of a quantum mechanical system. In numerical implementation of the Eq(2) the 3N dimensional Brownian motion is replaced by 3N independent, properly scaled one dimensional random walks as follows. For a given time t and integer n and l define [16] the vector in $R^{3N}$.

$$W(l) \equiv W(t,n,l) = (w_1^1(t,n,l), w_2^1(t,n,l), w_3^1(t,n,l), \dots\dots\dots\dots w_1^N(t,n,l), w_2^N(t,n,l), w_3^N(t,n,l)$$

where $w_j^i(t,n,l) = \sum_{k=1}^{l} \frac{\epsilon_{jk}^i}{\sqrt{n}}$

with $w_j^i(t,n,l) = 0$ for $i = 1,2,\dots\dots.N; j = 1,2,3 \; and \; l = 1,2,\dots.nt$.Here $\epsilon$ is chosen independently and randomly with probability $P$ for all $i,j,k$ such that $P = \left(\epsilon_{jk}^i = 1\right) = P\left(\epsilon_{jk}^i = -1\right) = \frac{1}{2}$. It is known by an invariance principle [17] that for every $\nu$ and $W(l)$ defined above .

$$\lim_{n\to\infty} P(\frac{1}{n}\sum_{l=1}^{nt} V(W(l))) \leq \nu$$

$$= P(\int_0^t V\big(X(s)\big)ds \leq \nu$$

Consequently for large n,

$$P[exp(-\frac{1}{n}(\int_0^t V\big(X(s)\big)ds) \leq \nu]$$

$$\approx P[\exp(-\frac{1}{n}\sum_{l=1}^{nt} V(W(l))) \leq \nu]$$

By generating $N_{rep}$ independent replications $Z_1, Z_2, \dots. Z_{N_{rep}}$ of

$$Z_m = \exp(-\frac{1}{n}\sum_{l=1}^{nt} V(W(l)))$$

And using the law of large numbers, $(Z_1 + Z_2 + \cdots + Z_{N_{rep}}) \big/ N_{rep} = Z(t)$ is an approximation to Eq(2)

Here $W^m(l), m = 1,2 \dots. N_{rep}$ denotes the $m^{th}$realization of $W(l)$out of $N_{rep}$ independently run simulations. In the limit of large t and $N_{rep}$ this approximation approaches an equality and forms the basis of a computational scheme for the solution of a many particle system with a prescribed symmetry.

### 2.2.2 Unit Analysis:

We show the unit analysis in the case of an attractive Gaussian potential as an example as follows:
The Hamiltonian for Rb gas with a symmetric trapping potential and for a general two body potential can be written as

$$[-\hbar^2/2m \sum_{i=1}^{N} \nabla_i'^2 + \sum_{i,j} V(r_{ij}') + \frac{m\omega^2}{2}(r'^2) = E\psi(r')$$

For a choice of attractive Gaussian potential as the two body potential

$V_\delta(r') = -g\delta_\sigma(r')$ with normalized Gaussian

$$\delta_\sigma(r') = \frac{\exp\left(-\frac{r'^2}{2\sigma^2}\right)}{\sqrt{2\pi}\sigma}$$

which tends to $\delta$ as $\sigma \to 0$ in the distribution sense and where $g$ can be calculated as follows

$$1 = g\int \; e^{-r^2/2\sigma^2} d^3r \qquad \text{yielding } g = \frac{1}{\sqrt{2\pi}\sigma}$$

the above equation can be written as

$$[-\hbar^2/2m \sum_{i=1}^{N} \nabla_i'^2 - \frac{\exp\left(-\frac{r'^2}{2\sigma'^2}\right)}{\sqrt{2\pi}\sigma'} + \frac{m\omega^2}{2}(r'^2) = E\psi(r')$$

The above Hamiltonian can be rescaled by substituting $\vec{r}' = s\vec{r}$ , $E = \lambda U$ and $\sigma' = s\sigma$

$$[-\frac{\hbar^2}{2ms^2} \sum_{i=1}^{N} \nabla_i^2 - \frac{s}{\sqrt{2\pi}\sigma} \exp\left(\frac{-r^2}{2\sigma^2}\right) + \frac{ms^2\omega^2 r^2}{2}]\,\psi(r) = \lambda U\psi(r)$$

Type equation here.

dividing the above equation by $-\frac{\hbar^2}{ms^2}$ throughout we get

$$[\frac{1}{2} \sum_{i=1}^{N} \nabla_i^2 + \frac{s}{\sqrt{2\pi}\sigma} \frac{ms^2}{\hbar^2} \exp\left(\frac{-r^2}{2\sigma^2}\right) - \frac{m^2 s^4 \omega^2 r^2}{2}]\,\psi(r) = -\frac{ms^2}{\hbar^2}\; \lambda U\psi(r)$$

Now let $\frac{m^2 s^4 \omega^2}{2} = 1 \Rightarrow s^2 = \frac{\hbar}{m\omega}$ is the natural unit of length. Let $\frac{ms^2 U}{\hbar^2} = 1 \Rightarrow U = \frac{\hbar^2}{ms^2} = \hbar\omega$ is the natural unit of energy. Then in the dimensionless form the above equation can be written as

$$[\frac{1}{2}\sum_{i=1}^{N}\nabla_i^2 + \frac{s}{\sqrt{2\pi}\sigma} e^{\frac{-r^2}{2\sigma^2}} - \frac{r^2}{2}]\psi(r) = -\lambda\psi(r)$$

**2.3. Precision of the numerical calculations:**

Next we calculate ground state density of trapped non-interacting BEC . To achieve that we choose $V_R = V_{int} = 0$ at $t = t_0$. Underneath we provide the results for this simplest case.

**2.1 Precision of the data associated with the calculations of Ground state density of non-interacting BEC in 1d with number of paths($N_{rep}$) = 100000, n=900 and t=1**

| x | Exact solution | FK solution | Exact density | FK density (this work) | error |
|---|---|---|---|---|---|
| 0 | 0.80501 | 0.80508 | 0.64804 | 0.64815 | -0.00011 |
| .1 | 0.80195 | 0.80167 | 0.64312 | 0.64268 | 0.00044 |
| .6 | 0.70189 | 0.70019 | 0.49264 | 0.49027 | 0.00237 |
| 1.1 | 0.50780 | 0.50566 | 0.25786 | 0.25569 | 0.00217 |
| 1.6 | 0.30369 | 0.30200 | 0.09220 | 0.09120 | 0.001 |

For more details in numerical analysis, one should take a look at one of our previous papers [18]

The above data can be represented pictorially as follows:

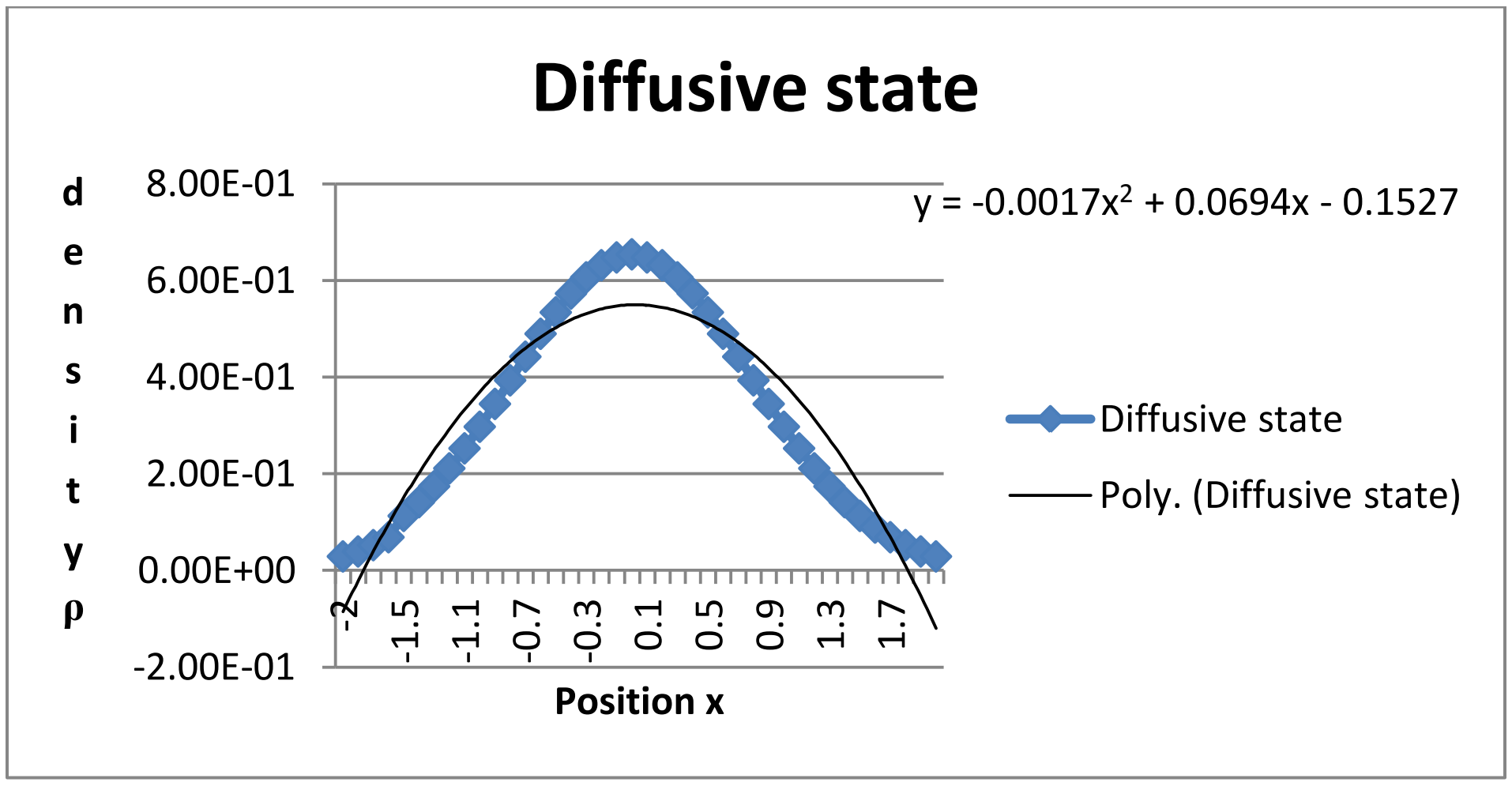

Fig 1 Ground state density of non interacting BEC in a trap potential in 1d

**3 Visualization of Anderson localization in quasiperiodic vs random system.**

**3.1. Anderson localization in bichromatic lattice.**

According to scaling theory of localization there is always localization in 1d. The interference effects on a short length scale are responsible for the fluctuations in the amplitude of the eigenfunction and reduction of the conductivity. The localization problem in 1d systems has been studied using various methods [19-25]. Exact analytical results and very accurate numerical data are available concerning the energy and disorder dependence of the localization length and probability distribution of various quantities of interest in 1d. Underneath we first describe the effect of quasi-periodic and random potential on a Bose gas and then compare the localization in two cases. Anderson localization has been observed for quasi-periodic optical lattices which realizes the so called A-A model. The A-A model is worth investigating **o**wing to their interesting feature of showing a transition from extended to expontially localized states when the disorder strength exceeds a critical value.

We study localization in a bichromatic lattice as it experimentally realizes Aubry-André (A-A) or Harper model and is given by the following Hamiltonian.

$$H = J\sum_m (|w_m\rangle\langle w_{m+1}| + |w_{m+1}\rangle\langle w_m|) + \Delta\sum_m \cos(2\pi\beta + \phi)|w_m\rangle\langle w_m|$$

where $|w_m\rangle$ is the Wannier state localized at the lattice site m, J is the side to side tunneling energy, and $\Delta$ is the strength of disorder, $\beta = k_2 / k_1$ is the ratio of two lattice wave numbers and $\phi$ is an arbitrary phase. Since in the tight binding approximation the bichromatic potential can be mapped to A-A model we study the full Hamiltonian

$$H = -\Delta_x / 2 + V(x)$$

We consider the motion of 100 trapped particles (($Rb^{87}$) $atoms$ ) in the following bi-chromatic potential (ref to Fig 2) $V_b = s_1 E_{R_1} Sin^2(k_1 x) + s_2 E_{R_2} Sin^2(k_2 x)$ .

Hence $V(x) = \frac{m}{2}\omega_x^2 x^2 + s_1 E_{R_1} Sin^2(k_1 x) + s_2 E_{R_2} Sin^2(k_2 x)$

Underneath we use the experimental parameters from the reference [26]

$k_i = 2\pi/\lambda_i \quad E_{R_i} = \frac{h^2}{2\pi\lambda_i^2}$

$s_1 = 10$ and $s_2 = 2 \quad \lambda_1 = 830.7\, nm \quad \lambda_2 = 1076.8\, nm$ , $\beta = 0.771452$

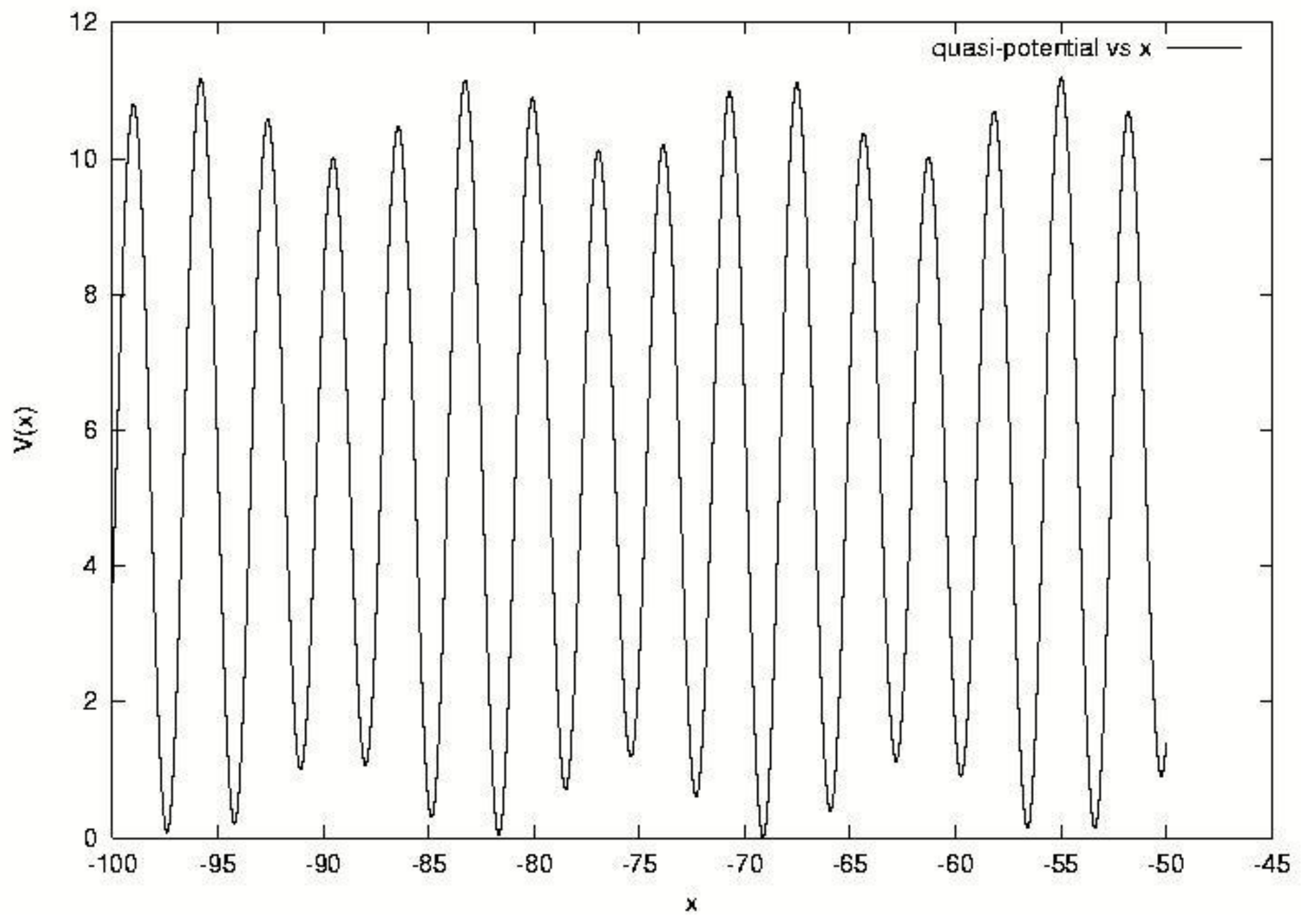

Fig 2 Plot of bi-chromatic potential $s_1 = 10$ and $s_2 = 2$ $\lambda_1 = 830.7\ nm$ $\lambda_2 = 1076.8\ nm$

In what follows in Fig 3 is the plot of density vs position curve for bichromatic disorder potential in 1d .

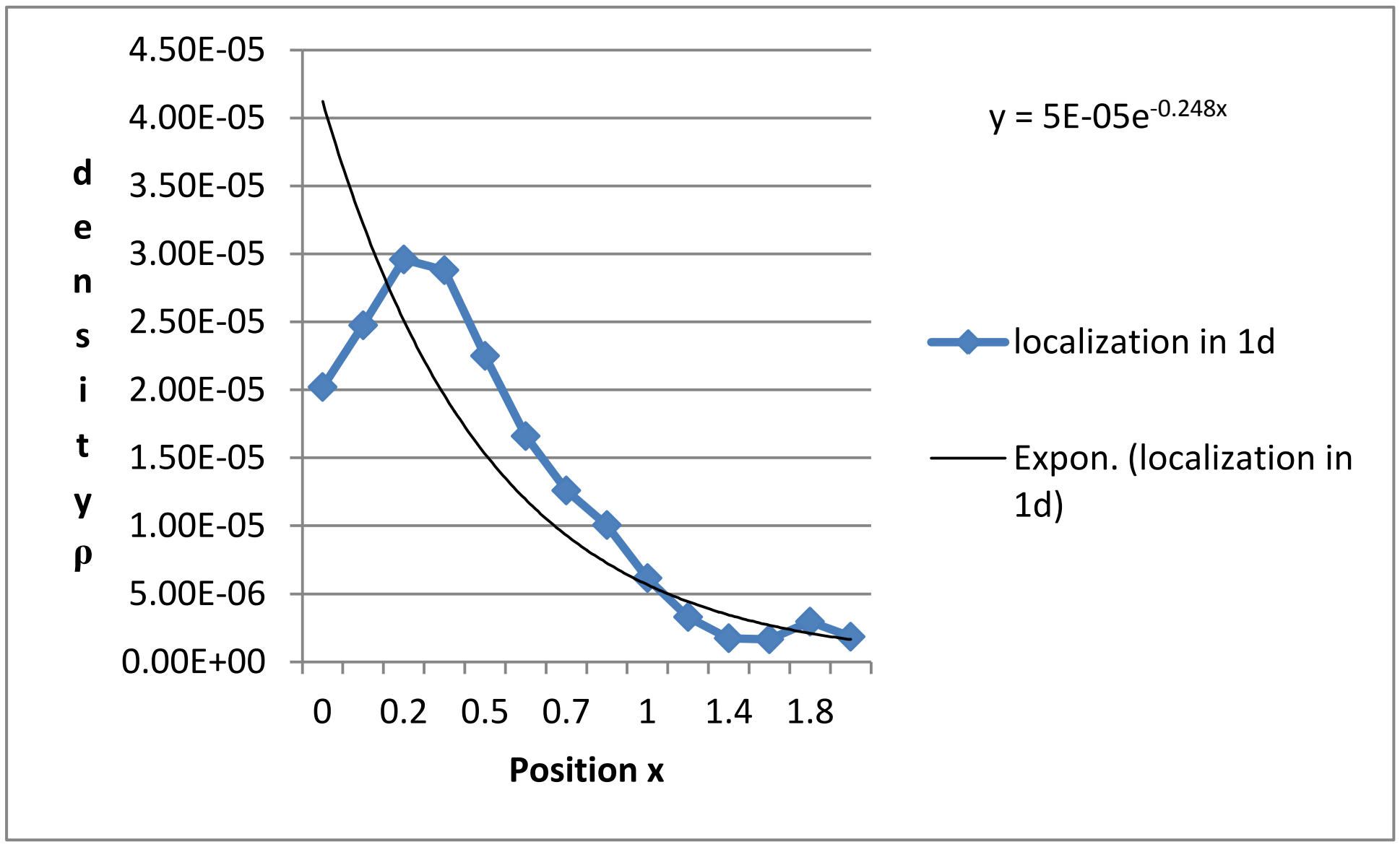

Fig 3. Anderson localization in 1d with quasi periodic potential

From exponential fit it turns out that the localization length is 4

### 3.2 Localization in 1d in presence of random disorder

Next we consider the effect of random potential in $Rb^{87}$ BEC in 1d. Let us first consider the Hamiltonian .

$H = -\Delta/2 + V_\delta(\vec{r})$

It should be noted that the three dimensional delta function above can be represented as $V_\delta(\overrightarrow{r}) = -g\delta(\overrightarrow{r})$ where $g$ is the strength of the interaction. This is a zero range interaction and this limit can be obtained from a finite range potential where the range approaches zero and strength is appropriately adjusted. For the finite –range interaction we use a Gaussian interaction $V_G(\overrightarrow{r}) = -ge^{-r^2/2\sigma^2}$ [27]. Or in other words in performing the actual simulation we represent Dirac delta function as a limiting case of Gaussian interaction as

$$\lim_{\sigma \to 0} V_G(\overrightarrow{r}) = V_\delta(\overrightarrow{r})$$

where $V_0$ can be calculated as follows

$$1 = g\int e^{-r^2/2\sigma^2} d^3r$$

yielding $g = \frac{1}{\sqrt{2\pi}\sigma}$

Now the one dimensional case

$H = -\Delta/2 - g\delta(x - x_j)$ has been studied before by Gavish et al[28]

Let us consider the Hamiltonian in 1d

$H = -\Delta/2 - g\delta(x - x_j) + \frac{1}{2}x^2$

At t=0 the trapping potential is suddenly switched off.

That is for t>0

$V(x) = -g\delta(x - x_j)$

Underneath in Fig (4) and (5) we show the localization in the presence of random potential with and without the trapping potential respectively. In Fig 6 and 7 density vs position plots are being shown for different values of interaction strength ‘g’.

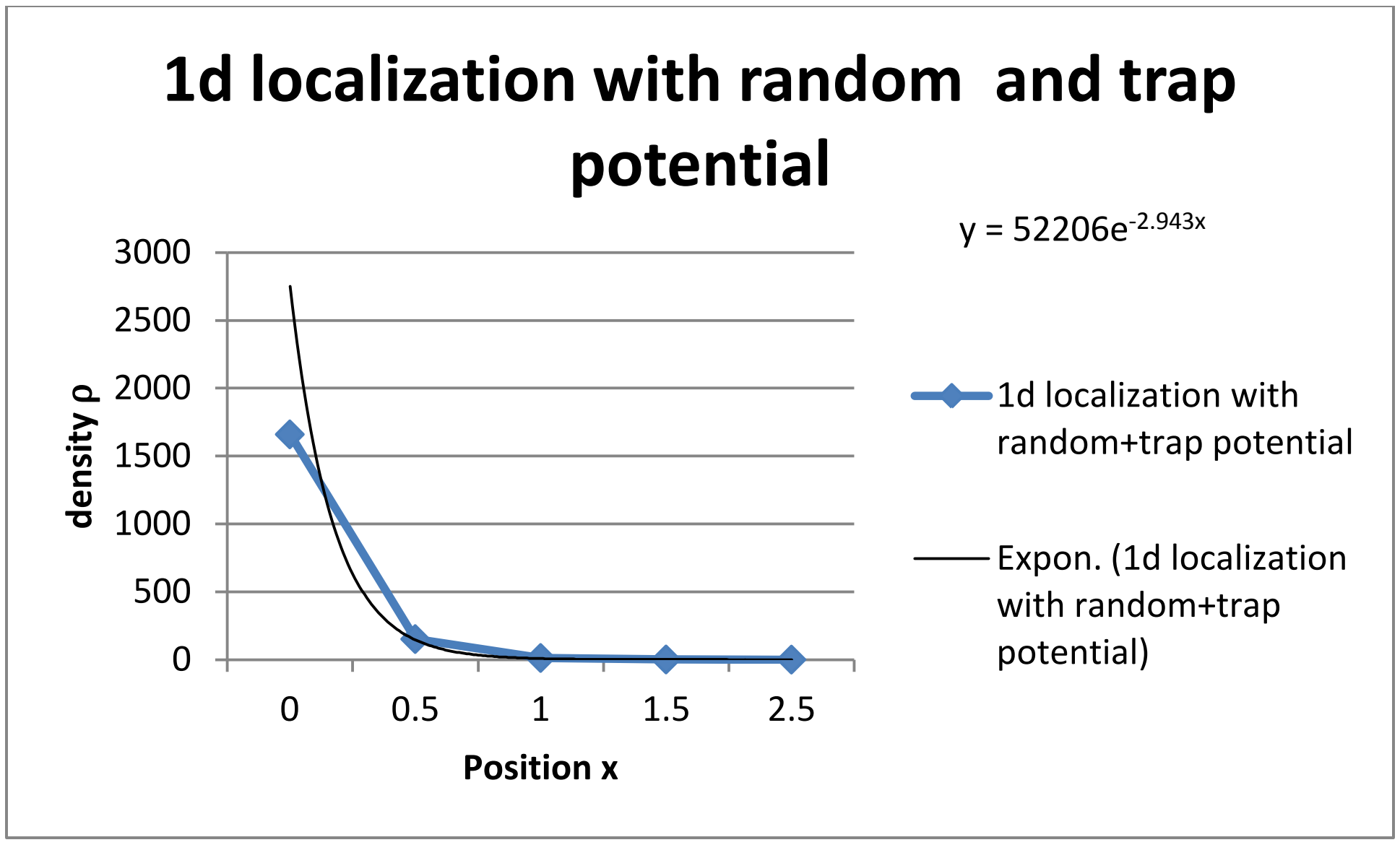

Fig 4.  Anderson localization in 1d with true random potential with trapping potential

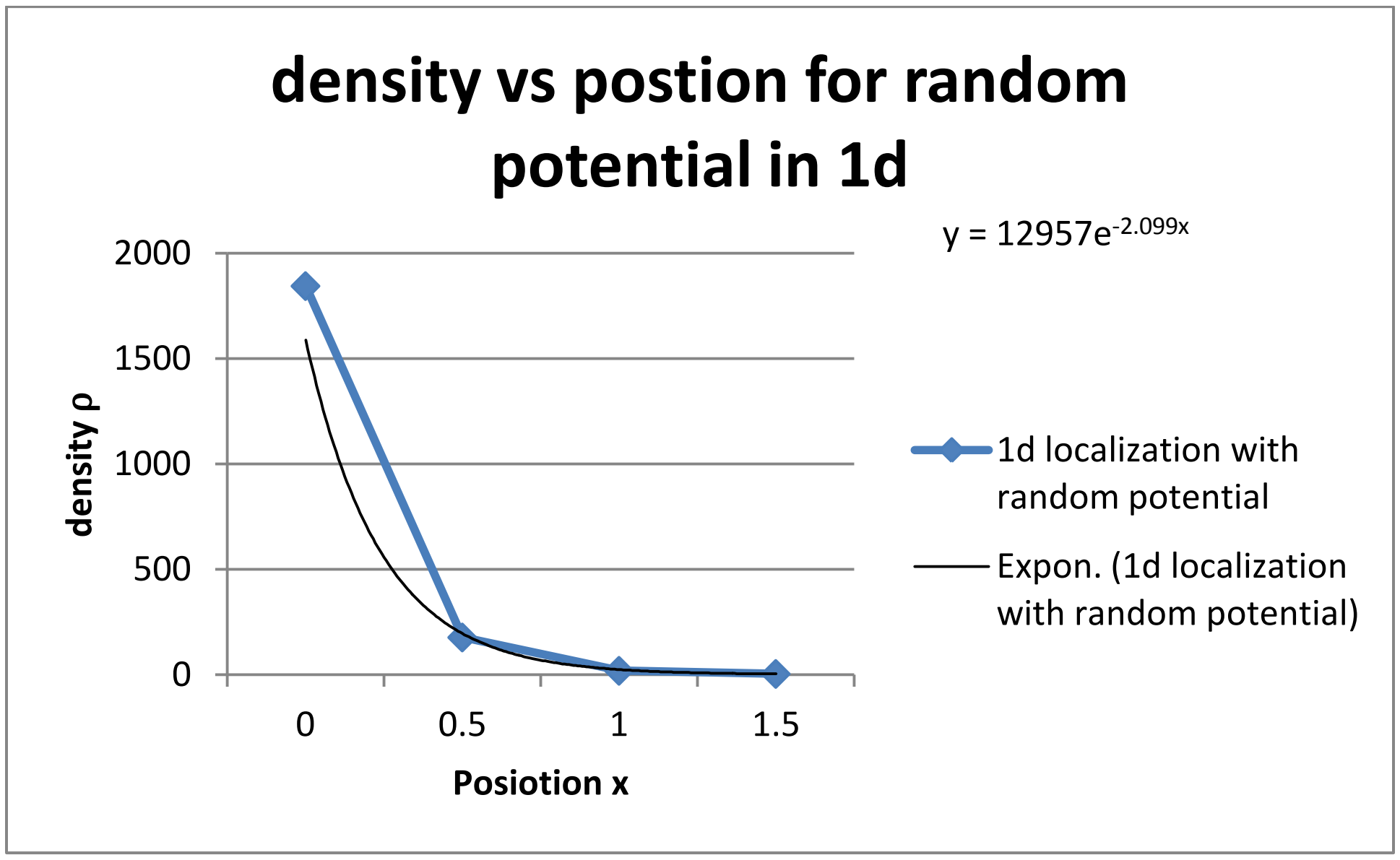

Fig 5.  Anderson localization in 1d with true random potential without the trapping potential

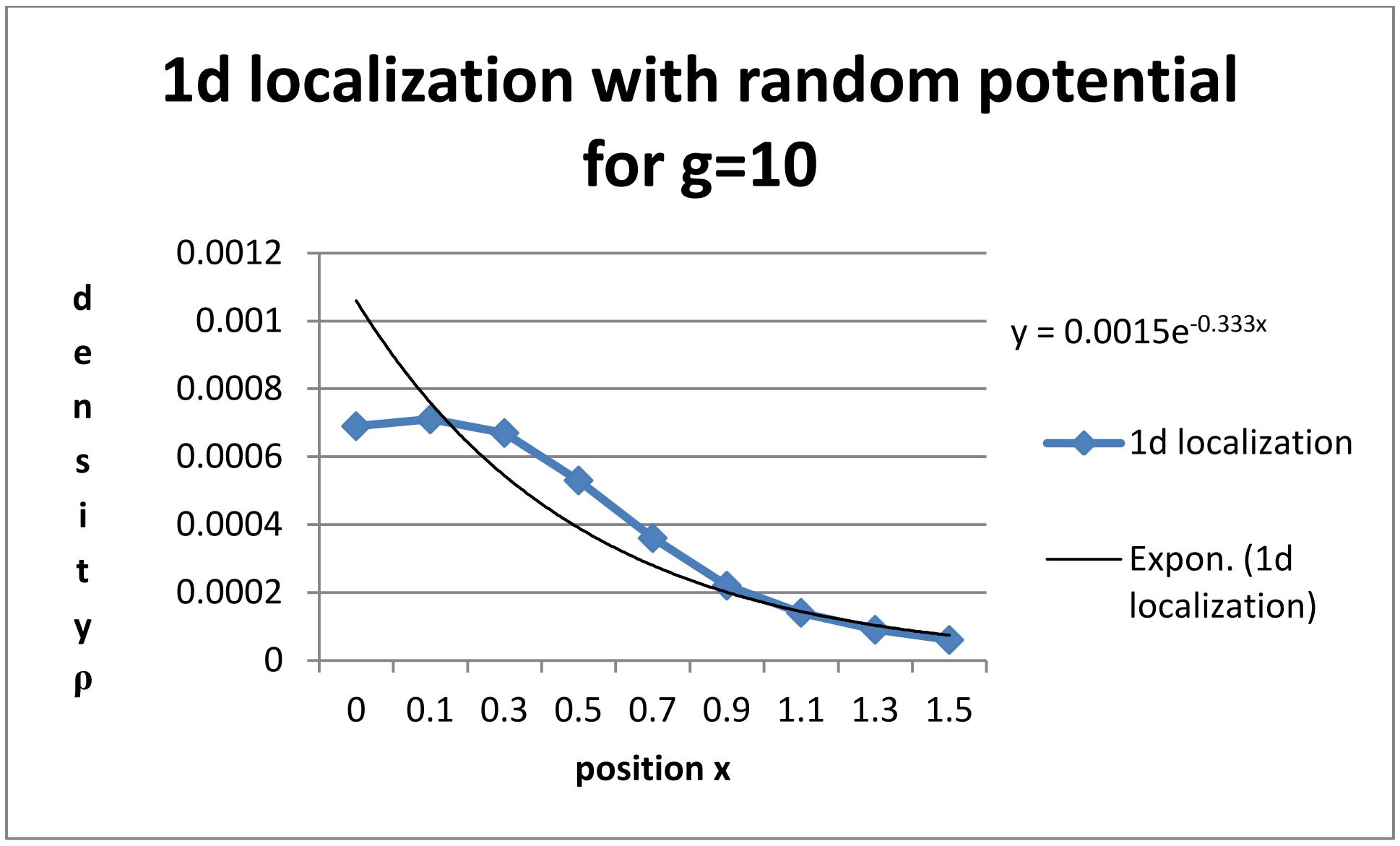

Fig 6. Anderson localization in 1d with true random potential with disorder strength g=10

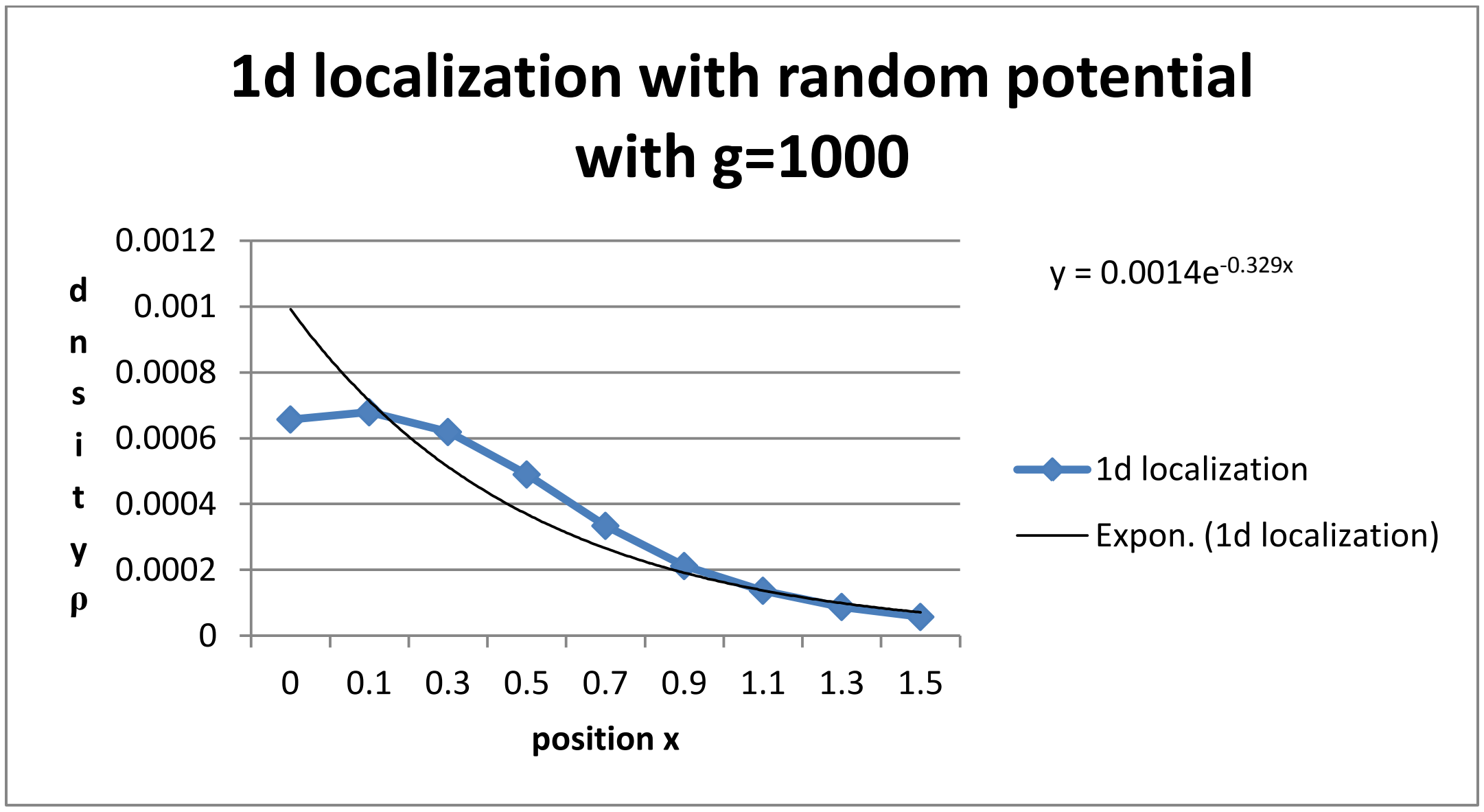

Fig 7 Anderson localization in 1d with true random potential with disorder strength g=1000

In Fig 3 we see exponential localization and as we used the experimental parameters of [26], it corresponds to the threshold of $\Delta / J = 6$ i.e., the localization occurs only when $\Delta / J$ exceeds the critical value of 6. But in Fig 4-7, we see that localization persists for any value of disorder. The localization length corresponding to Fig 6 and 7 are 3.03 and 3.12 and almost independent of the value of the disorder present.

According to the Anderson theory, particle wave will exponentially localize as a result of multiple scattering from the disorder potential and will be characterized by an exponential decay at large distances $L_{loc}$ and $\gamma = 1/L_{loc}$ is called the Lyapunov exponent and $L_{loc}$ is called the localization length. To probe Anderson localization in a system one can calculate the Lyanunov exponent either analytically [29] or numerically. However calculating Lyapunov exponents for a system may not be very easy all the time. So as an alternative we calculate **Imaginary time mean square displacement**: which seems to be an elegant way to distinguish between extended and localized states and can be found by looking at their time evolution[30].More precisely one can calculate the mean square distance $r^2(t)$ travelled by the particle up to time t:

$$r^2(t) = \sum |x|^2 \, |exp[itH]\varphi_0(x)|^2$$

where $\varphi_0$ is the initial wave packet and $exp[itH]$ is the evolution operator. In the case of perfect periodic crystal or without any disorder in the system $r^2(t)$ increases in time t, while in the case of localization $r^2(t)$is bounded uniformly with time. We calculate this quantity in all the following cases to probe for localization.

**3.1 Variation of Imaginary Time Mean square displacement with time t for BEC in 1d without the random potential with number of paths $N_{rep} = 1$, n=900 and t=8-48 with reference to Fig 1**

Imaginary Time Mean square displacement $r^2(t) = \sum |x|^2 \, |exp[itH]\varphi_0(x)|^2$

| t | mean square distance |
|---|---|
| 8 | 62561.594101 |
| 16 | 191812.368405 |
| 24 | 438610.445992 |
| 32 | 1230319.373950 |
| 40 | 2323076.110784 |
| 48 | 3697358.142701 |

We observe that that in the absence of random potential the Imaginary Time Mean Square displacement increases with time. We also find that if the random potential is included in the form of a 1d Gaussian potential the mean square distance is bounded by $1.02509x10^{-3}$.

**3.3 Localization in the BEC in 3d with one dimensional random potential.**

Now suppose $V_R = 0$ at $t = t_0$. Then following reference [31] the Schrödinger equation for the above Bose condensate reads as the following where $\mu_c$ is the chemical potential of the Bose condensate

$$[-\Delta/2 + V_{int} + V_{trap}]\psi_c(\vec{r},t) = \mu_c \psi_c(\vec{r},t)$$

where $V_{trap} = \frac{1}{2}\sum_{i}^{N}[x_i{}^2 + y_i{}^2 + z_i{}^2]$

In 3d as soon as the $V_R$ is turned on for $t > t_0$ a new energy scale is created at the mobility edge $\varepsilon_c$ :

$H = -\Delta/_2 + V_{trap} + V_R$ where $V_R = -g\delta(\overrightarrow{r})$

and underneath we have generated two plots (Fig 8(a) and Fig 8(b)) using the data obtained from the above Hamiltonian.

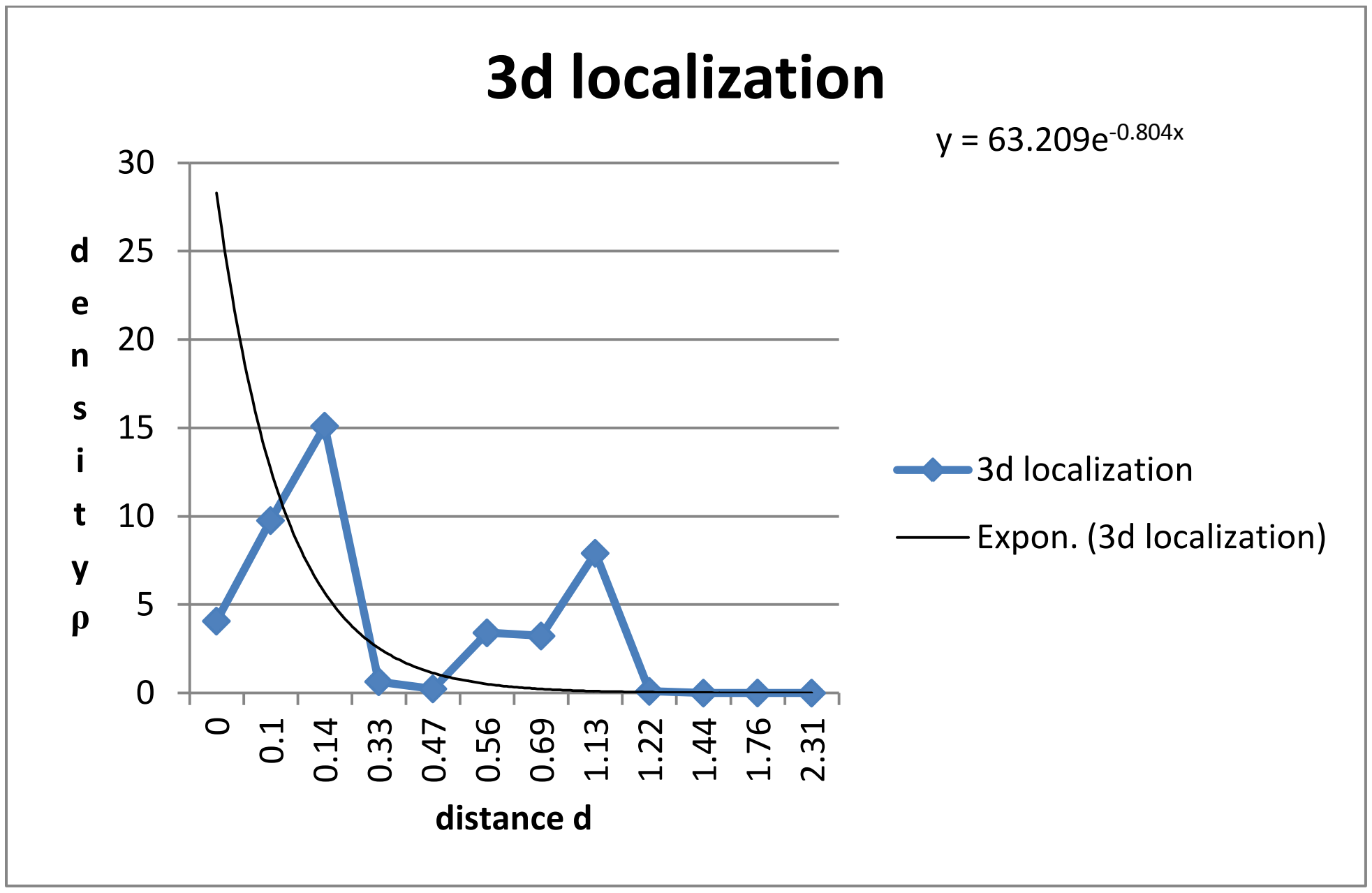

Fig 8(a) localization in 3d with random potential; plot of density vs distance ‘d’

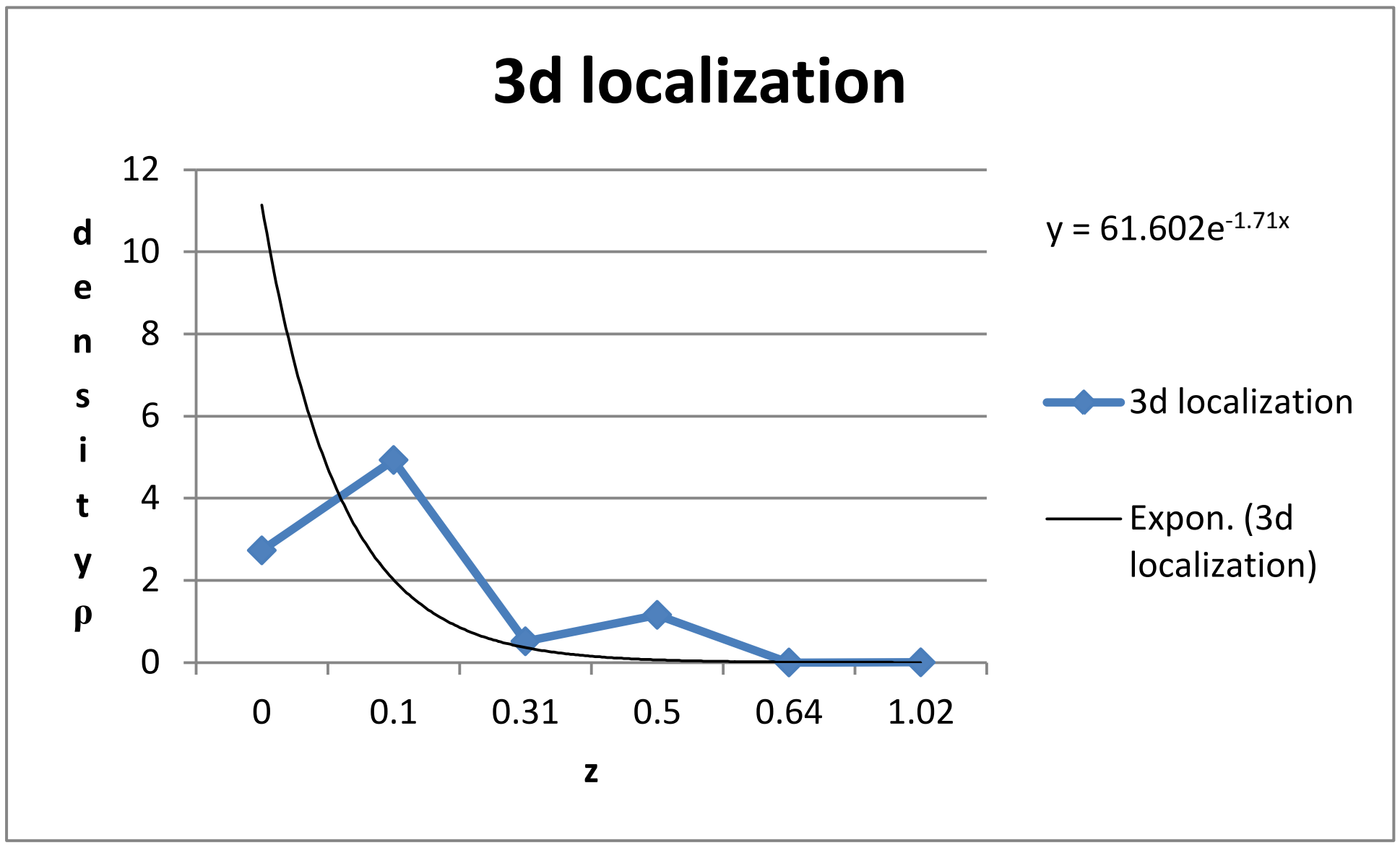

Fig 8(b) localization in 3d with random potential; plot of density vs z

**3.3 Calculation of the mobility edge:**

In order to decide whether or not a state is localized it is often sufficient to consider the second moment $P^{-1} = \sum_r |\Psi(r)|^4$, where $\Psi(r)$ is the eigenvector of the corresponding symmetry state with eigenvalue $\lambda$ . The above quantity is called the inverse participation number (ipn) and it is nonzero if it falls on the interval of energy comprising localized states (pure point spectrum)[32] and vanishes for extended states(absolutely continuous spectrum). In Fig 9 we plot Inverse Participation Number (ipn) with energy and observe that the critical energy after which the spectrum is continuous is 0.0729. This critical energy is the mobility edge for $Rb^{87}$ . To calculate energy '$\lambda$'of the condensate in 3d we employ Eq(3) of section 2.

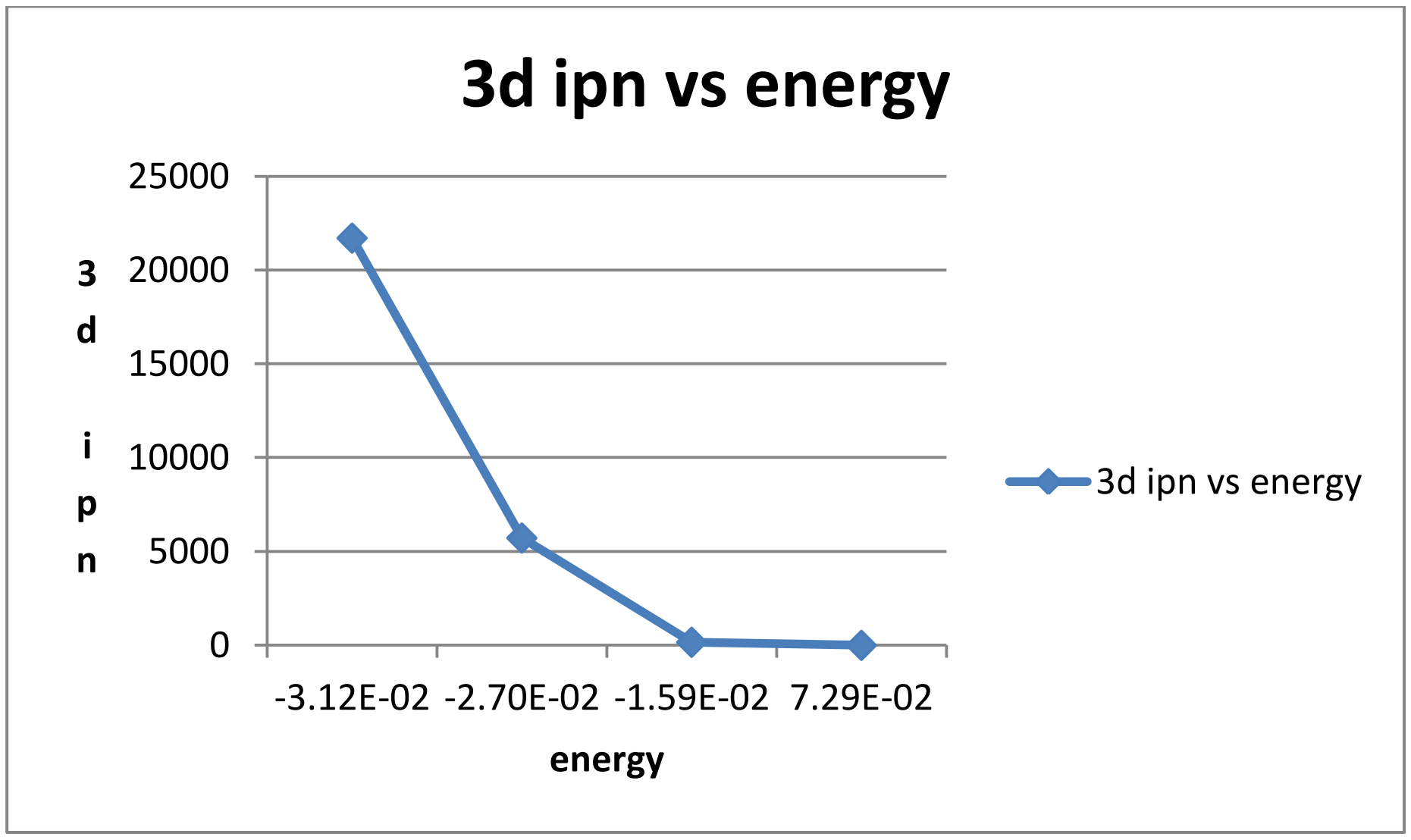

Fig 9 ipn vs energy in 3d

## 4 Delocalization in presence of an interaction

It is well-known that Anderson localization is the effect of constructive interference which results from the overall coherence of the atomic cloud. Now if anytime a decoherence is introduced in the system, it destroys the constructive interference and delocalization takes place. Sometimes this delocalization can be induced in the system using a suitable interaction. Underneath we give an account of such incident in the case of bichromatic potential in 1d and true random potential in 3d.

### 4.1 Delocalization in presence of interaction in the case of quasi-periodic potential.

Fig 10(a) shows the ground state of BEC. Fig 10(c)-Fig 10(e) show the effect of Morse interaction on the bi- chromatic potential. As the interaction strength 'g' increases the density profile becomes more and more coherent. As a result of introducing quasi-periodic disorder the periodicity of is partly broken as shown in the Fig 10(b) and the spectrum consists of both extended and localized states. By adding Morse interaction in the Hamiltonian the spectrum becomes more and more coherent (Fig 10(c)- Fig 10(d)). As a matter of fact the periodicity of Fig 10(a) is again restored in Fig 10(c) once a small amount of interaction g was incorporated in the quasi-random Hamiltonian. In Fig 10(e) we see that the effect of disorder is completely delocalized by the Morse interaction and we see a transition from Anderson glass to the coherent BEC again. Similar analysis has been done in ref [33]

$$[-\Delta/2 + V_{int} + V_{trap} + V_b]\psi_c(\vec{r},t) = \mu_c \psi_c(\vec{r},t)$$

Where $V_{trap} = \frac{1}{2}\sum_{i}^{N}[x_i^{\,2} + y_i^{\,2} + z_i^{\,2}]$

$$V_{int} = V_{Morse} = \sum_{i,j} V(r_{ij}) = \sum_{i<j} g[e^{-\alpha(r-r_0)}(e^{-\alpha(r-r_0)} - 2)]$$

$$V_b = s_1 E_{R_1} Sin^2(k_1 x) + s_2 E_{R_2} Sin^2(k_2 x)$$

$k_i = {}^{2\pi}/_{\lambda_i}$ $\quad E_{R_i} = \frac{h^2}{2\pi\lambda_i^2}$

$s_1 = 10$ and $s_2 = 2$ $\quad \lambda_1 = 830.7\,nm$ $\quad \lambda_2 = 1076.8\,nm$

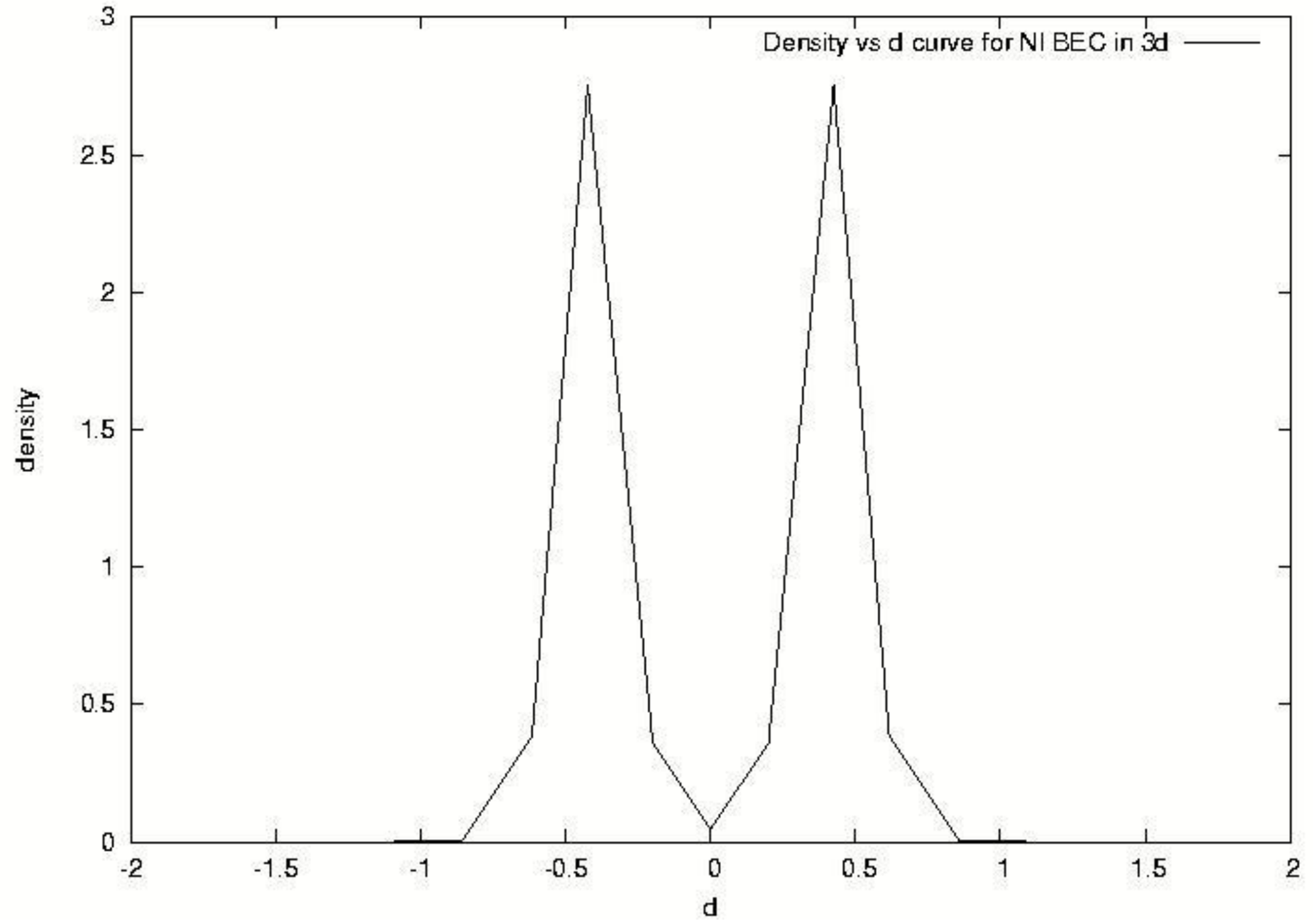

Fig 10(a) Density vs d curve for non-interacting BEC in 3d

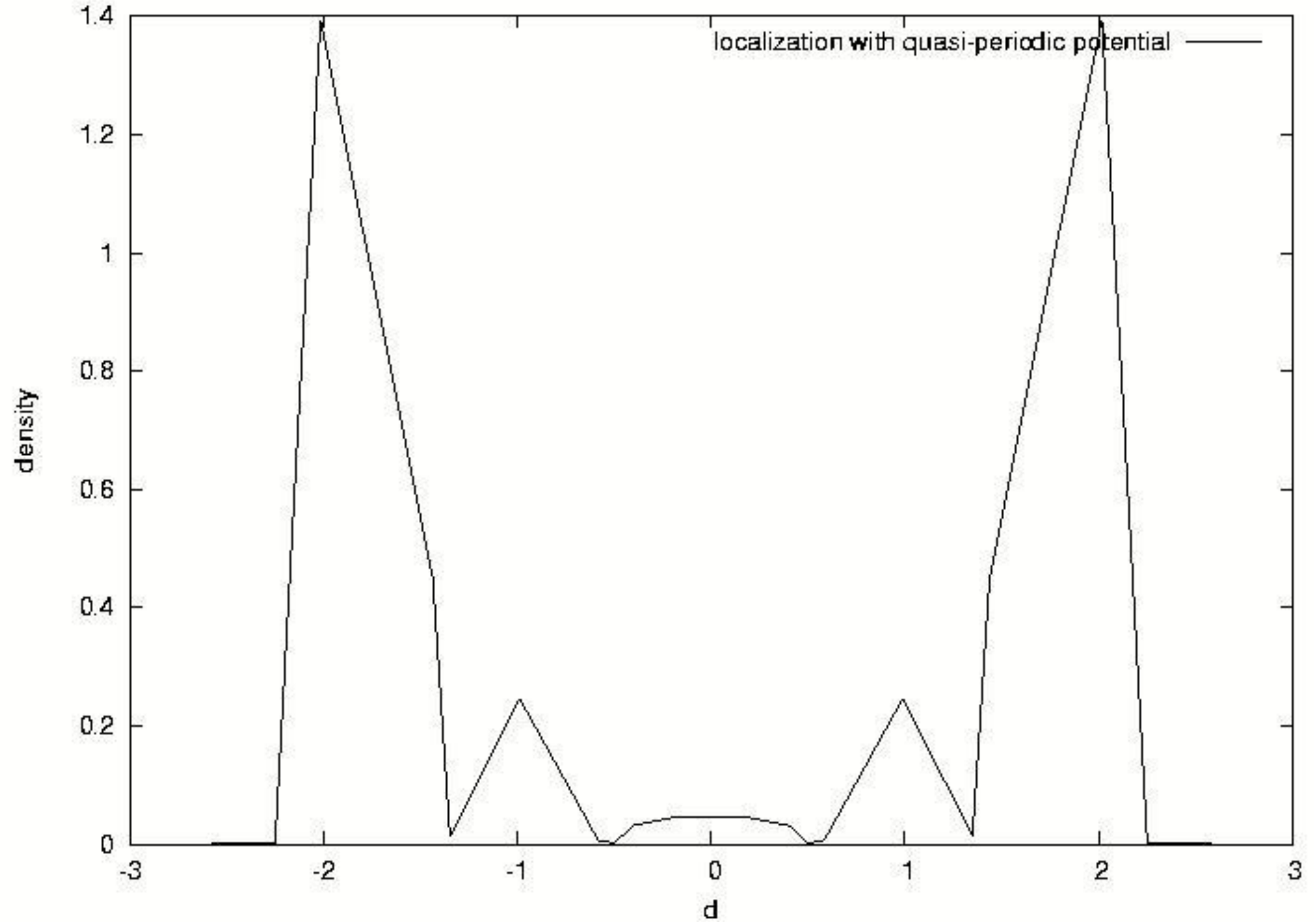

Fig 10(b) localized and extended states in 3d with quasi potential

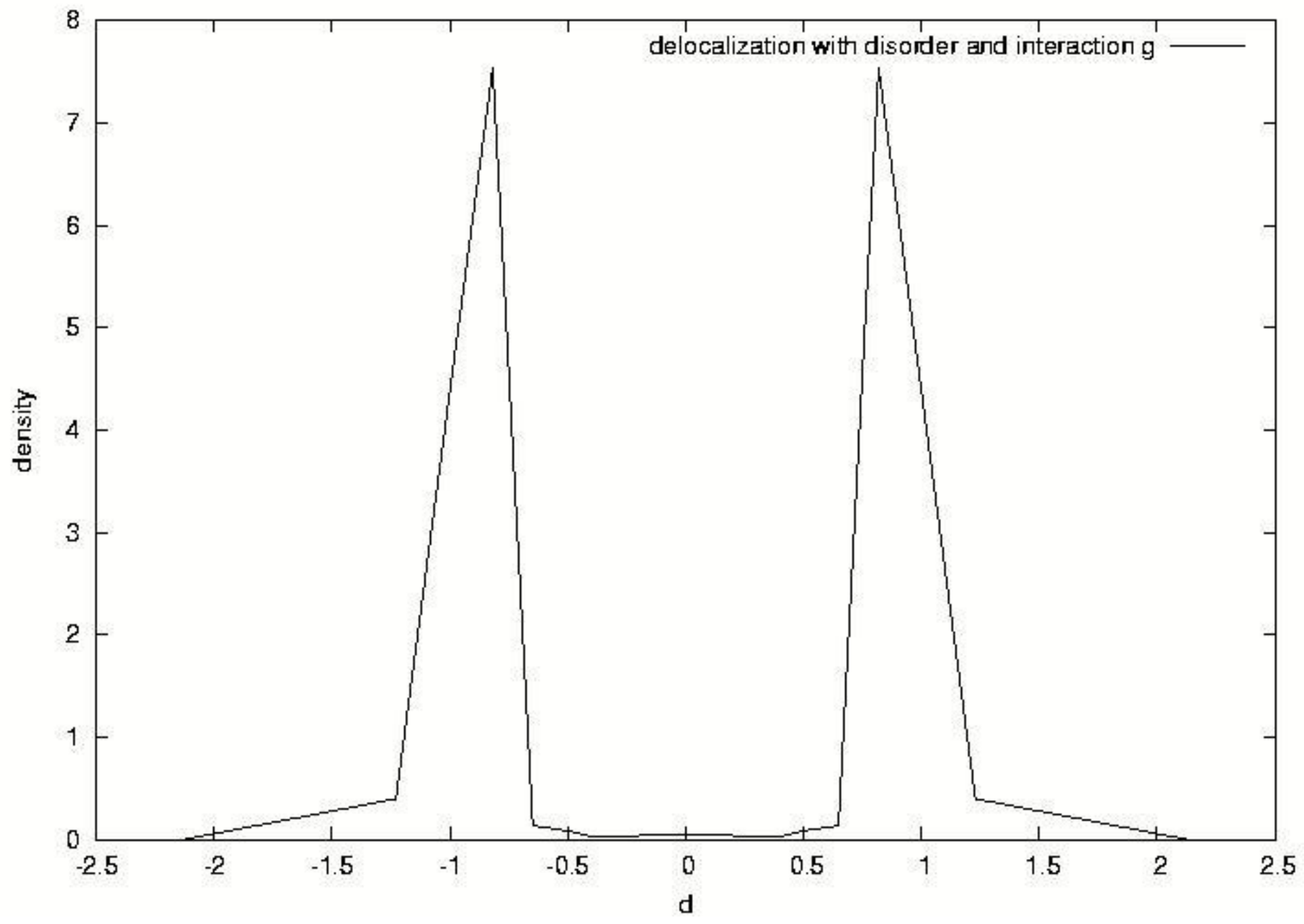

Fig 10(c) delocalization in 3d with disorder and interaction g

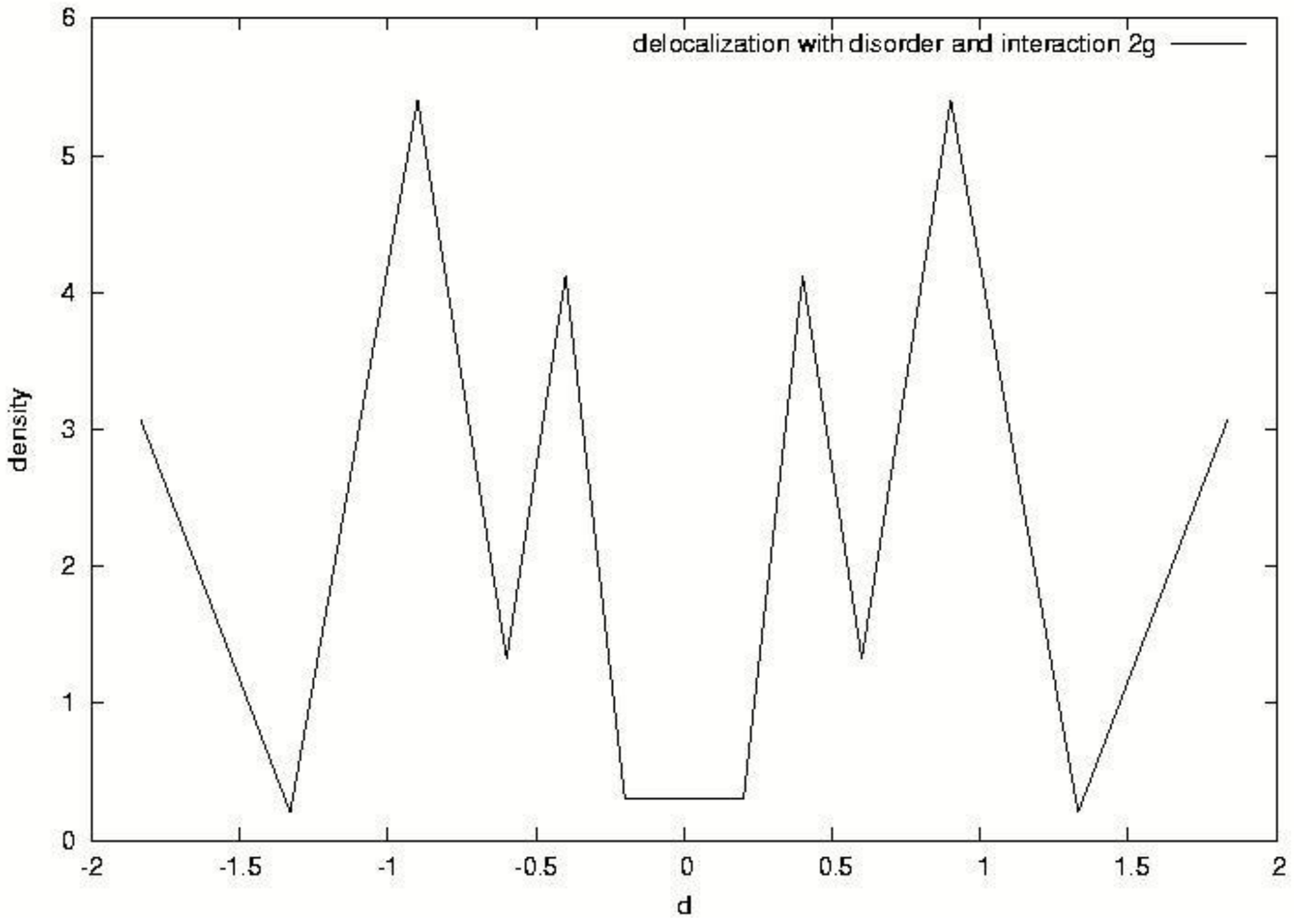

Fig 10(d) delocalization in 3d with disorder and interaction 2g

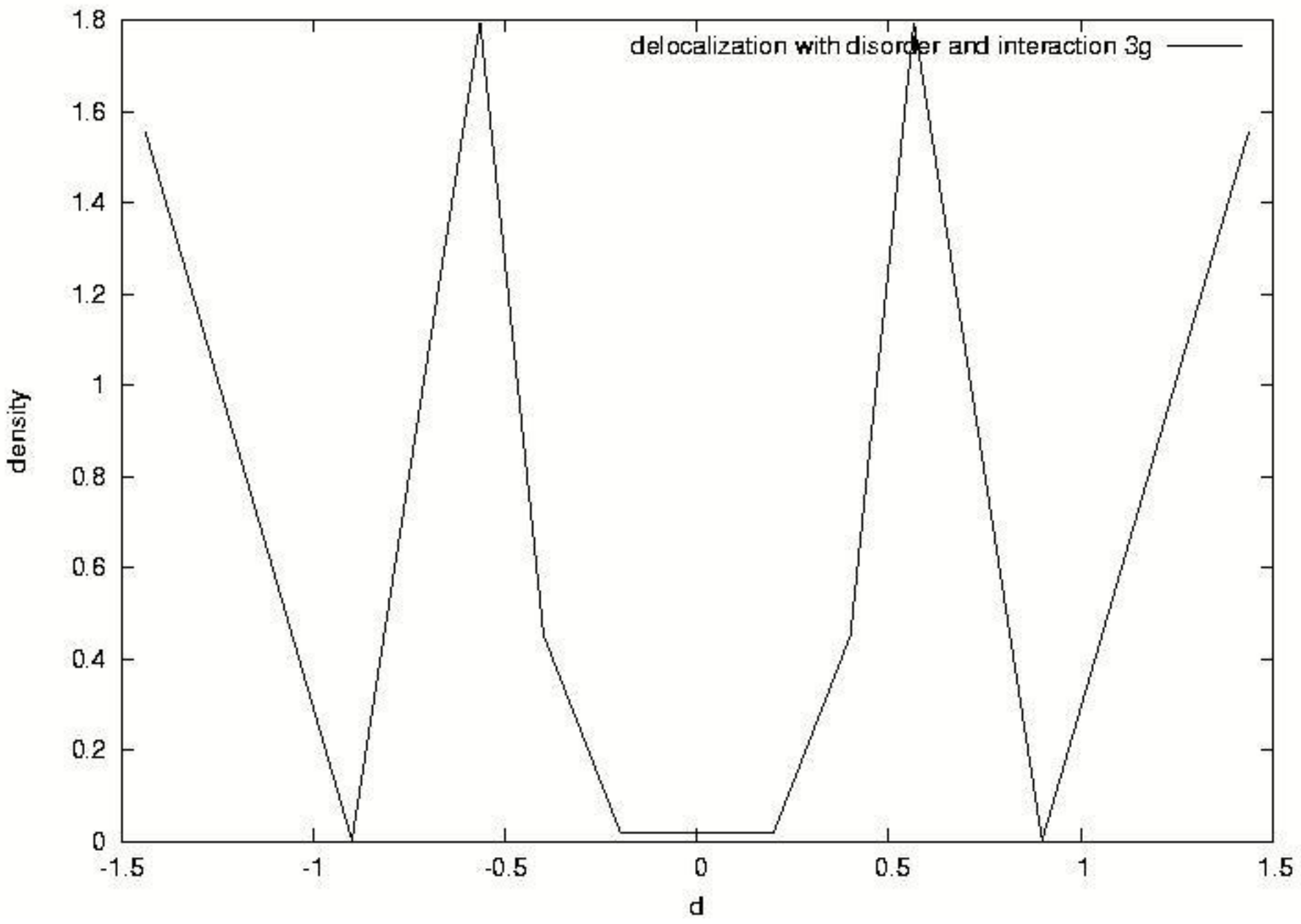

Fig 10(e) complete delocalization in 3d with disorder and interaction 3g

**4.2 Delocalization in presence of interaction in the case of speckle potential.**

We also consider the motion of particles ( $87_{Rb}\ atoms$ ) in the speckle potential. To visualize the speckle potential we adopt the same analytic form of attractive sech-squared-shape potential as used in the ref [34-35] because the potential well can act as a reflector for the multiple interference of the condensate.This multiple interference explains the non-trivial localization effects in the presence of random disorder. In fig 11(a) we observe the localization effect due to the above speckle potential and in Fig 11(b) and Fig 11(c) we see the effect of delocalization due to the Morse potential .

$V(z) = -b * sech^2(\frac{z}{\beta})$ where b and $\beta$ are the parameters involved in the model.

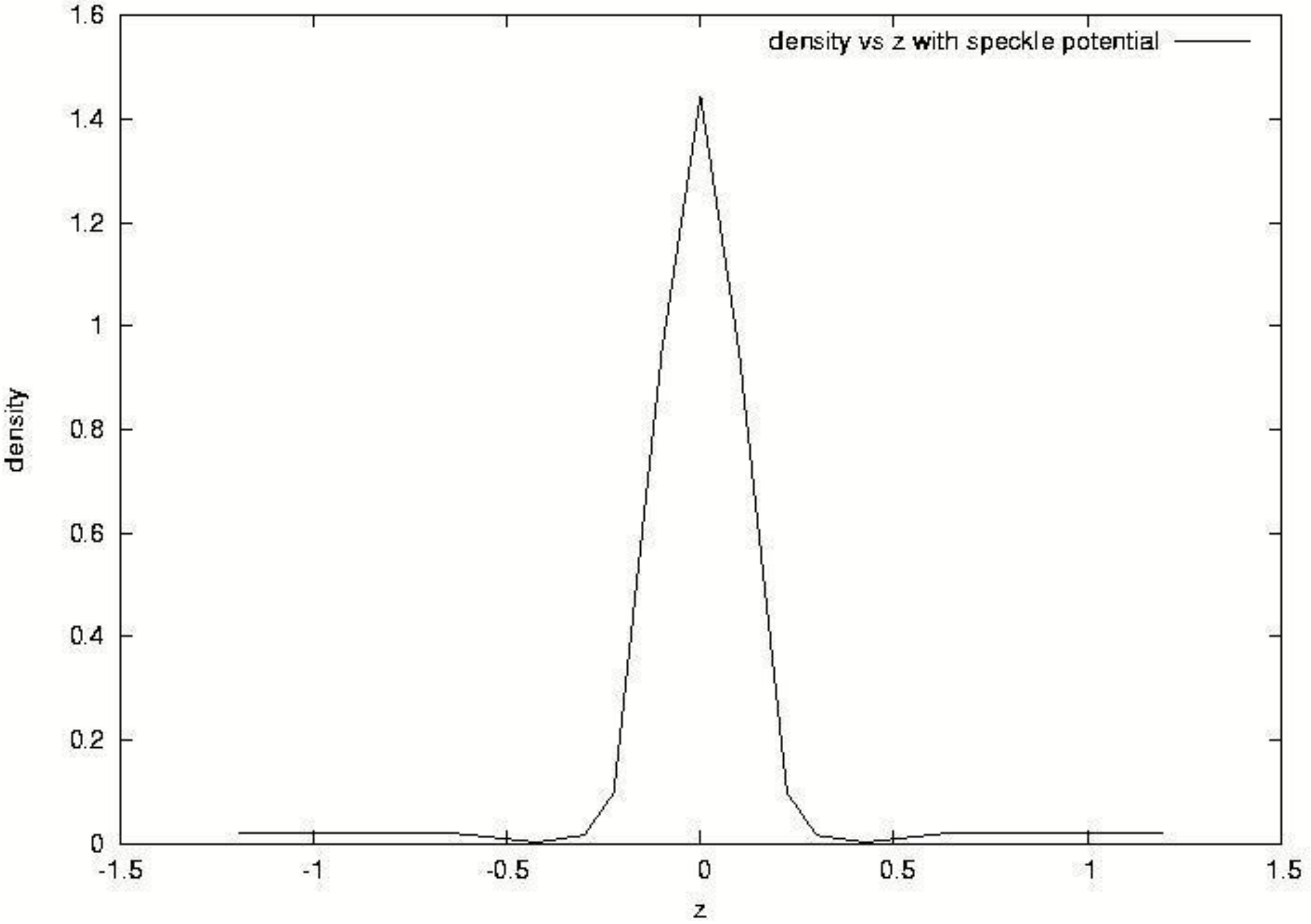

Fig 11(a) Localization in presence of speckle potential

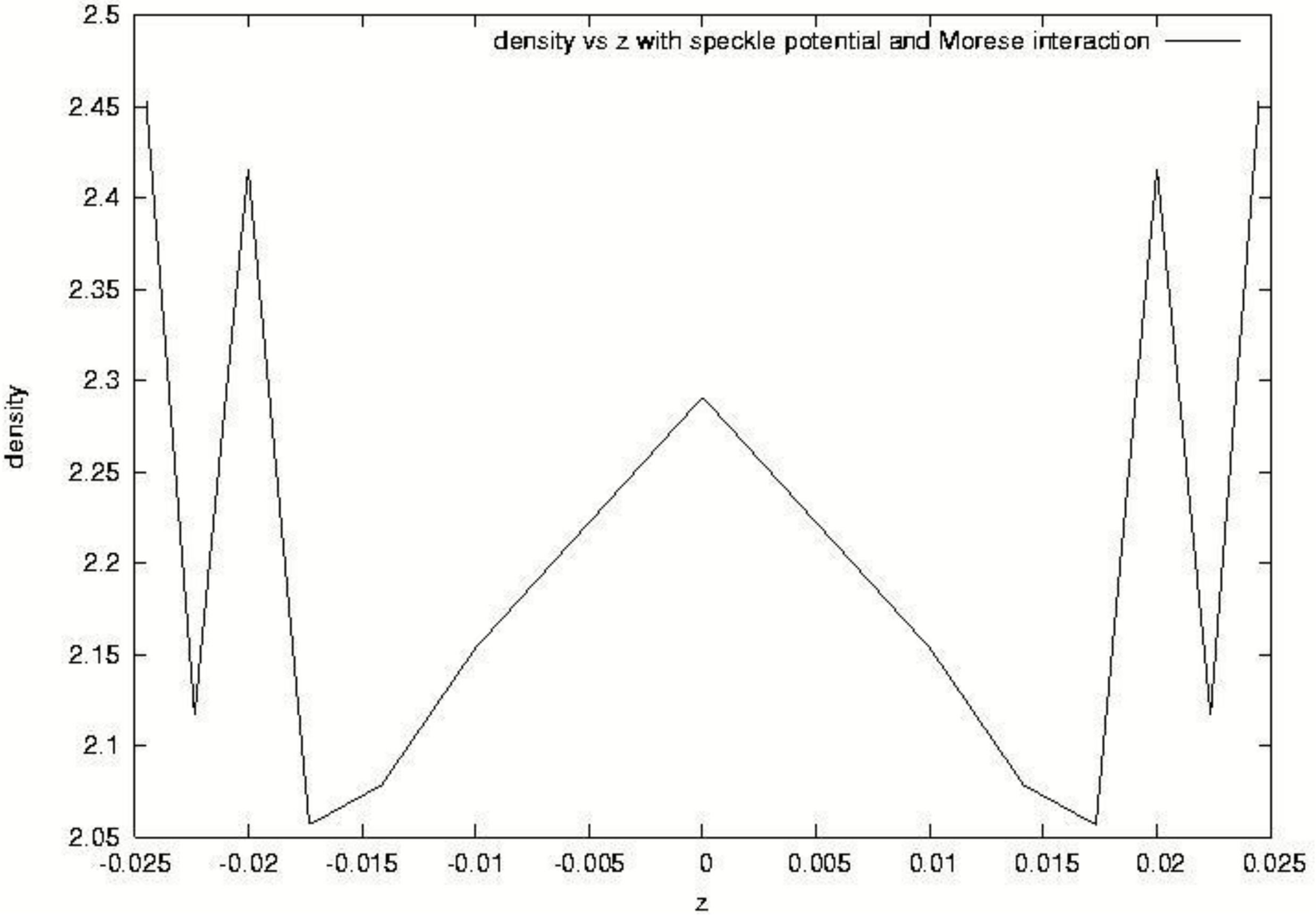

Fig 11(b) delocalization with speckle potential and Morse interaction

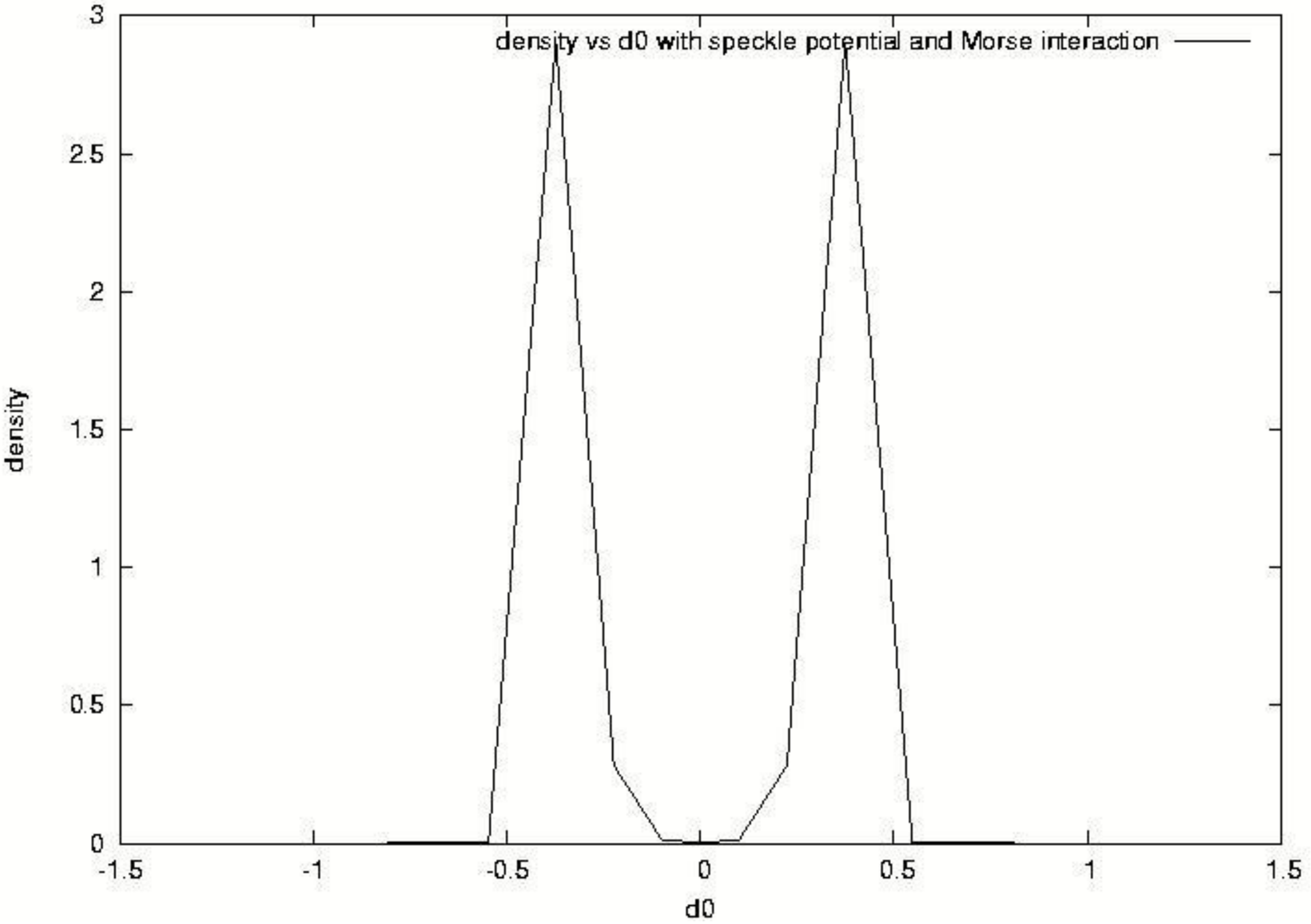

Fig 11(c) Complete delocalization

**5 Discussions:** From our observation we find that always in 1d and sometimes in 3d the wave functions in random potentials are localized. The incommensurate case is intermediate and is sometimes localized and sometimes extended. This establishes the fact there exists a mobility edge which separates the point and extended spectrum. We visualize that as bi-chromatic potential can induce localization in a system of cold atoms , it is worth using this potential as an effective tool for implementing quasi-disorder in the case of an actual experiment. Both in the case of bichromatic potential in 1d and a real 1d disorder potential we observe a localization i.e., an exponential decay of the wavefunction. But in the case of bi-chromatic potential it occurs particularly for $\Delta/J = 6$ whereas in the case of a true random potential in 1d the localization persists for any value of disorder. We also observe that interaction delocalizes the effect of disorder in Bose system in 3d.In ref [33] we see that for $K^{39}$ interaction induces a delocalization and there is a transition from Anderson glass to coherent BEC via small fragmented BEC whereas in the similar situation in our simulation for $Rb^{87}$ we have a direct transition from Anderson glass to coherent BEC without any intermediate state. We verify for each localized state the mean square distance is bounded and for the diffusive state the mean square distance grows with time. The other work in 3d which are worth mentioning can be found in the literature [36-38]. In the future we would like to study the density of states and critical exponents for these systems. Currently we are also trying to simulate the recent experiment [4,5] on the non-interacting spin-polarized gas of $Rb^{87}$and $K^{40}$ atoms to observe the three-dimensional Anderson Localization of ultracold matter.

**Acknowledgements:** This work was supported by the Department of Science and Technology, India(award no SR/WOS A/PS-32/2009)